\documentclass[aps,prl,twocolumn,showpacs,amsmath]{revtex4}
 \usepackage[colorlinks=true,citecolor=blue]{hyperref}
 \usepackage{graphics} 
 \usepackage{epsfig}
   \begin{document} 
\title{
Two-particle non-local Aharonov-Bohm effect from two single-particle emitters 
} 
\author{
Janine Splettstoesser$^{1}$, 
Michael Moskalets$^{1,2}$, 
and Markus B\"uttiker$^{1}$
}
\affiliation{
$^1$D\'epartement de Physique Th\'eorique, Universit\'e de Gen\`eve, CH-1211 Gen\`eve 4, Switzerland \\
$^2$Department of Metal and Semic. Physics, NTU "Kharkiv Polytechnic Institute", 61002 Kharkiv, Ukraine
}
\date\today
\begin{abstract}
We propose a mesoscopic circuit in the quantum Hall effect regime comprising two uncorrelated single-particle sources and two distant Mach-Zehnder interferometers with magnetic fluxes, which allows in a controllable way to  produce orbitally entangled electrons. 
Two-particle correlations appear as a consequence of erasing of which path information due to collisions taking place at distant interferometers and in general at different times.  
The two-particle correlations manifest themselves as an Aharonov-Bohm effect in noise while the current is insensitive to magnetic fluxes.
In an appropriate time interval the concurrence reaches a maximum and a Bell inequality is violated.
\end{abstract}
\maketitle

\begin{figure}[t] \includegraphics[width=3.3in]{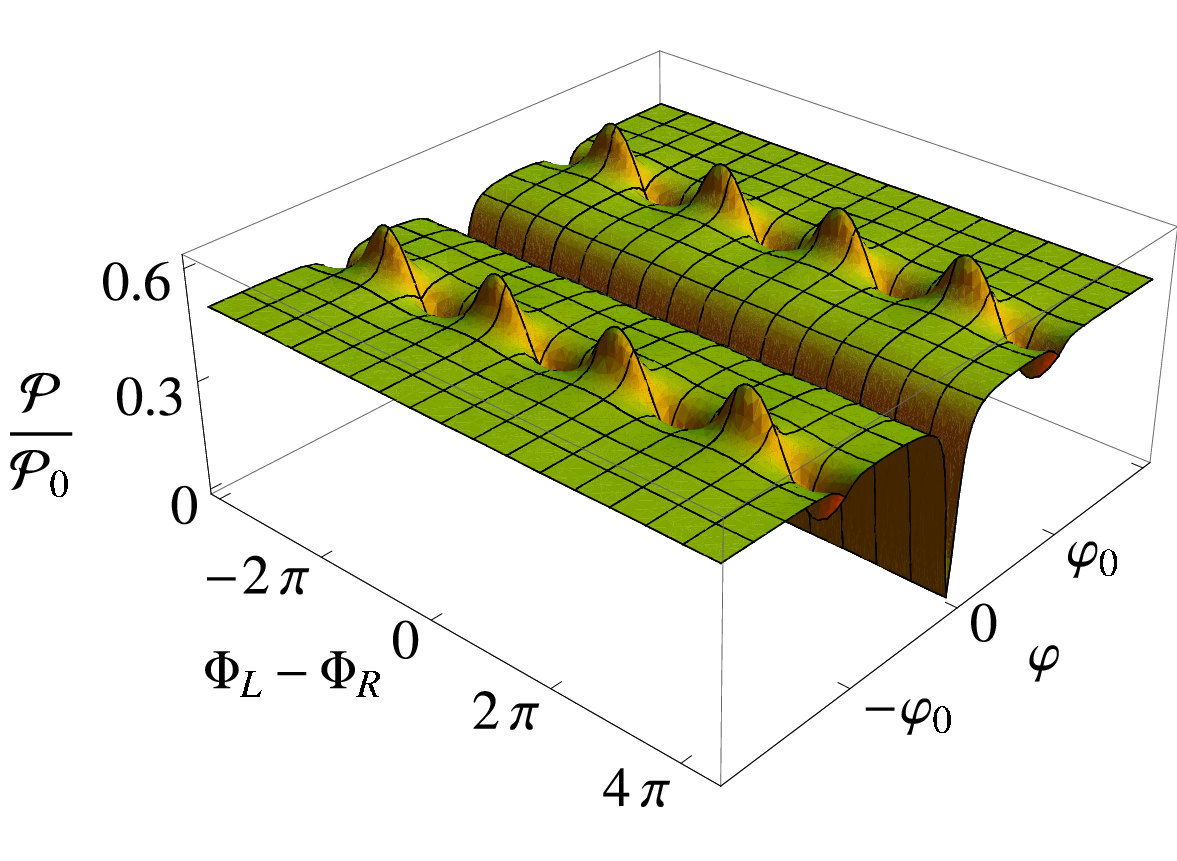}
\caption{
Two-particle Aharonov-Bohm oscillations in the shot noise correlation ${\cal P}_{12}$ as a function of the difference in magnetic fluxes $\Phi_A - \Phi_B$ and difference of phase $\varphi$ of the potentials $U_{A}(t)$ and $U_{B}(t)$.
}\label{fig_noise}\end{figure}

\textit{Introduction.}---
Interference phenomena are the most prominent feature of quantum mechanics. 
Of particular interest are interference effects in multi-particle states.
For example in optics, the Hanbury Brown-Twiss effect \cite{hanburybrown56} and the Hong-Ou-Mandel (HOM) effect \cite{hong87} both result from  two-particle interference of photons emitted by two independent sources.
In mesoscopics, electrons can play a role similar to photons in optics.
In an electrical circuit with currents incoming from different (uncorrelated)
equilibrium contacts the noise can show interference even if the currents exhibit no interference contribution \cite{buttiker91}. 
\\ \indent
Recently a single-particle emitter \cite{feve07} was experimentally realized on the basis of a quantum capacitor in a two-dimensional electron gas in the integer quantum Hall effect regime.
Subject to an appropriate large amplitude potential the capacitor emits a {\it single} electron during the first half-cycle and a {\it single} hole during the second half-cycle. 
With such an emitter it is possible to inject single electrons and holes in a non-equilibrium state into an electrical conductor.
Injected particles can be guided by edge states and encounter splitters realized by quantum point contacts (QPC). 
These states can be considered as an analogue of short photon pulses produced by a laser. 
By using two such sources and tuning the times when they emit particles one can force emitted particles to collide at some QPC.
Tuning can be achieved by varying the phase difference between the two potentials acting on the capacitors. 
Such a collision erases which-path information for particles leaving the QPC and it promotes the appearance of two-particle correlation effects.
Based on this simple idea an electronic analogue of an optical HOM interferometer was suggested \cite{olkhovskaya08}.
This interferometer shows a noise suppression due to two-particle correlations arising locally 
when particles collide at a 
QPC. 
\\ \indent
\begin{figure}[b]
\includegraphics[width=3.in]{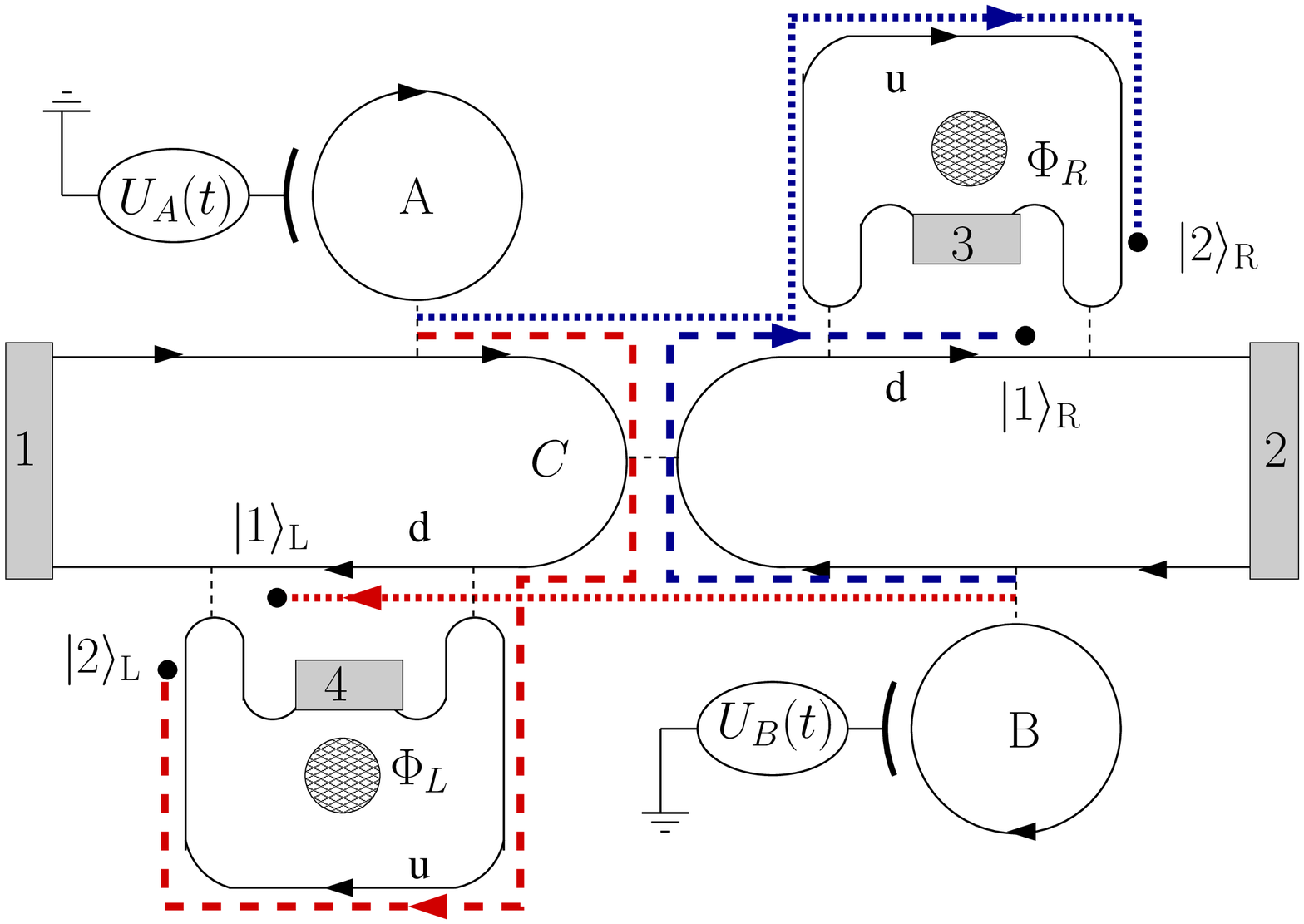}
\caption{(color online). Emitters $A$ and $B$ driven by potentials $U_A (t)$ and $U_B (t)$ inject single particles into edge states (solid lines). After scattering at a center QPC ($C$) particles reach Mach-Zehnder interferometers with Aharonov-Bohm fluxes $\Phi_\mathrm{L}$ and $\Phi_\mathrm{R}$. The colored (dotted and dashed) lines show possible two-particle amplitudes which lead to particle collisions erasing which path information.} \label{fig_collision}
\end{figure}
In
this Letter we 
propose a set-up, where two-particle correlations arise non-locally in time and in space.
This is particularly intriguing, when the currents are magnetic-flux independent, namely when the width of the electron pulses is small with respect to the arm length differences of the Mach-Zehnder interferometers (MZIs),
see Fig. \ref{fig_collision}.
Our geometry resembles the optical Franson interferometer \cite{franson89} but the underlying physics  is different.
The uncorrelated electrons (holes) emitted by the sources $A$ and $B$ and propagating along different arms of the same interferometer can collide at the interferometer exit if the times of emission, defined by the potentials $U_{A}(t)$ and $U_{B}(t)$, are properly chosen. 
If the difference of the arm lengths for both interferometers is the same then the electron collisions take place at both interferometers (possibly at different times).
This makes the two two-particle amplitudes (with one electron going through the left and another going through the right interferometers) indistinguishable hence interfering. 
Corresponding amplitudes are shown in Fig.\,\ref{fig_collision} in dotted and dashed lines. 
Such an interference results in two-particle correlations arising non-locally in space and time. 
As a direct manifestation of this interference, the noise shows an Aharonov-Bohm (AB) effect \cite{aharonov59} with the total flux determined by both the magnetic fluxes $\Phi_\mathrm{L}$ and $\Phi_\mathrm{R}$ threading the {\it distant} interferometers.
We analyze the two-electron state emitted by our set-up and find it to be fully entangled with collisions in place and completely disentangled in the case that no collisions are present. 
We arrive at the same conclusion by analyzing a Bell inequality \cite{Bell65}, which can be violated even at partial overlap of electron wave-packets. 
\\ \indent
The two-particle AB effect 
in electrical conductors was discussed theoretically \cite{samuelsson} and investigated experimentally \cite{neder07} in a Hanbury Brown-Twiss interferometer geometry \cite{yurke92}. 
The novelty of our proposal is the quantized injection of electrons and the possibility of controlling the appearance of two-particle  correlations.
In comparison to proposals discussed in the literature for a dynamical generation of entanglement \cite{entpump}, the present scheme deals with electron-electron (and hole-hole) rather then electron-hole entanglement. 
\\ \indent
\textit{Model and formalism.}--- 
The system is schematically shown in Fig. \ref{fig_collision}. Two mesoscopic capacitors, $A$ and $B$,  are contacted by QPCs, with reflection (transmission) coefficients $r_{A},\ (t_{A})$ and $r_{B}\ (t_{B})$, to chiral edge states. 
The mesoscopic capacitors are driven by time-dependent, homogeneously applied potentials $U_{A}(t)$ and $U_{B}(t)$, with equal frequency $\Omega$ and a large amplitude such that the capacitors serve as sources of single electrons and holes. 
The particles emitted are transmitted or reflected at the center QPC ($C$), with reflection (transmission) coefficients $r_\mathrm{C},\ (t_\mathrm{C})$. Before reaching the contacts $1$ to $4$ the signals traverse the lower (d) or upper (u) arms of two MZIs \cite{MZI,chung}, $L$ and $R$,  pierced by magnetic fluxes, $\Phi_\mathrm{L}$ and $\Phi_\mathrm{R}$.
The beam splitters of the MZI have the reflection (transmission) coefficients $r_\beta^\mathrm{l}$ ($t_\beta^\mathrm{l}$) and $r_\beta^\mathrm{r}$ ($t_\beta^\mathrm{r}$).
\\ \indent
We use a scattering matrix approach \cite{moskalets08,KSL08} and describe the mesoscopic capacitor by a Fabry-Perot like amplitude \cite{moskalets08}, 
$S_\alpha(t,E)= r_\alpha+t_\alpha^2\sum_{q=1}^\infty {r_\alpha}^{q-1}e^{iqE\tau_\alpha-i\Phi^q_\alpha(t)} $, $\alpha=\mathrm{A,B}$, 
depending on the energy of an incoming particle and the time it exits.  
Here $\tau_\alpha=h/\Delta_\alpha$ and $\Delta_\alpha$ is the capacitor's level spacing;
$\Phi^q_\alpha(t) =  \frac{e}{\hbar}\int_{t-q\tau_\alpha}^{t}dt'U_\alpha(t')$.
The scattering matrix of the full system also
depends on
the central QPC and the MZIs. 
A phase  $\Phi_{\beta\alpha}^{\mathrm{u(d)}}(E)=E\tau_{\beta\alpha}^{\mathrm{u(d)}}/ \hbar+\bar{\Phi}_\beta^{\mathrm{u(d)}}$ 
is accumulated when a particle coming from source $\alpha$ traverses the upper (lower) arm of the interferometer $\beta=\mathrm{L,R}$. The time for this traversal is 
$\tau_{\beta\alpha}^{\mathrm{u}}$ ($\tau_{\beta\alpha}^{\mathrm{d}}$) and the
phase $\bar{\Phi}_{\beta\alpha}^{\mathrm{u}}$ ($\bar{\Phi}_{\beta\alpha}^{\mathrm{d}}$) depends on the magnetic flux. 
We are interested in the case of slow driving, meaning that the frequency $\Omega$ is much smaller than 
the inverse of the lifetime of particles in the cavity, \textit{without requesting restrictions on $\Omega$ with respect to the time scales related to the entire system}. In order to calculate the currents into contact $1$ and $2$, we need to know the scattering matrix elements for scattering from contacts $1$ and $2$ into contacts $1$ and $2$. We find
\begin{eqnarray}
S_{11}(t,E) & = & r_\mathrm{C}\left[t_\mathrm{L}^\mathrm{l}t_\mathrm{L}^\mathrm{r}S_{A}(t-\tau^\mathrm{u}_{\mathrm{L}A},E)e^{i\Phi_{\mathrm{L}A}^\mathrm{u}(E)}\right.\nonumber\\
&& \left.+r_\mathrm{L}^\mathrm{l}r_\mathrm{L}^\mathrm{r}S_{A}(t-\tau^\mathrm{d}_{\mathrm{L}A},E)e^{i\Phi_{\mathrm{L}A}^\mathrm{d}(E)}\right].
\end{eqnarray}
The elements $S_{12}(t,E)$, $S_{21}(t,E)$ and $S_{22}(t,E)$ are found analogously,  while the elements which describe the scattering from contact $3$ or $4$ are time-independent.
\\ \indent
\textit{The current.}---
A periodic potential with period $\mathcal{T}=2\pi/\Omega$, results in a current at contact $j$, \cite{splett08}
\begin{eqnarray}
\label{eq_current_general} 
I_j(t)& = & \frac{e}{h}\int
dE\sum_{n=-\infty}^{\infty}\sum_{l=1}^2\left[f(E) - f(E+\hbar n\Omega)\right]\\
&& 
\times\int_0^\mathcal{T} \frac{dt'}{\mathcal{T}} e^{in\Omega(t-t')} S^{}_{jl}(t,E) S_{jl}^*(t',E)\ .\nonumber
\end{eqnarray}
This equation is valid for finite frequency and 
arbitrary
amplitude driving. 
The current at  contact, say, $1$, 
$I_1(t)= R_{C}I_{1A}(t)+T _{C}I_{1B}(t)$ (here $T_{C}=1-R_{C}=|t_{C}|^2$), 
consists of a current coming from source $A$ ($B$) reflected (transmitted) at the central QPC and passed through the interferometer $\mathrm{L}$.
At zero temperature, the partial current $I_{1\alpha}(t)$, $\alpha = A, B$, comprises a classical part due to the current generated by the capacitor, $I_\alpha(t)=\frac{e^2}{2\pi i}\frac{\partial U_\alpha}{\partial t}S^{(0)*}_\alpha(t)\frac{\partial}{\partial E}S^{(0)}_\alpha(t)$, and an interference part,
\begin{eqnarray}\label{eq_single_particle}
&&I_{1\alpha}(t) = R_\mathrm{L}^\mathrm{l}R_\mathrm{L}^\mathrm{r} I_\alpha(t-\tau^\mathrm{d}_{\mathrm{L}\alpha})+T_\mathrm{L}^\mathrm{l}T_\mathrm{L}^\mathrm{r} I_\alpha(t-\tau^\mathrm{u}_{\mathrm{L}\alpha})\\
&&+\frac{e\gamma_\mathrm{L}/\pi} {\tau^\mathrm{u}_{\mathrm{L}\alpha}-\tau^\mathrm{d}_{\mathrm{L}\alpha}}
\mathrm{Im}\left\{S_\alpha^{(0)*}(t-\tau^\mathrm{u}_{\mathrm{L}\alpha})S_\alpha^{(0)}(t-\tau^\mathrm{d}_{\mathrm{L}\alpha})e^{-i\Phi_\mathrm{L}}\right\}.\nonumber
\end{eqnarray}
The current in contact $2$ is found analogously. Here we introduced the instantaneous scattering matrix of the cavity $\alpha$ at the Fermi energy $\mu$, $S_\alpha^{(0)}(t)=S_\alpha(\mu-eU_\alpha(t))$. The transmission probability at each MZI beam splitter is $T_{\beta}^j=1-R_{\beta}^j=|t_{\beta}^j|^2$ and $\gamma_\beta=(R^\mathrm{l}_{\beta}R^\mathrm{r}_{\beta}T^\mathrm{l}_{\beta}T^\mathrm{r}_{\beta})^{1/2}$ is a product of the reflection and transmission coefficients. Furthermore the flux enclosed by the interferometer $\beta$ is given by
$\Phi_\beta = \bar{\Phi}^\mathrm{u}_\beta - \bar{\Phi}^\mathrm{d}_\beta$,  in units of $\Phi_{0}/(2\pi)$, where $\Phi_{0} = h/e$ is the magnetic flux quantum. As the coherent emission of quantized charge attracts our special attention, we now treat the case of small transmission of the cavities' QPCs, where the current emitted by a cavity is a series of well separated pulses of opposite sign for the emission of electrons and holes. 
We resort to a description in the Breit-Wigner regime, and assume that one electron and one hole are emitted per period from cavity $\alpha$ at times $t^\mathrm{e}_\alpha$ and $t^\mathrm{h}_\alpha$, described by current pulses, having a Lorentzian shape with half width $\Gamma_\alpha$ \cite{splett08}.
\\ \indent
The last term in Eq. (\ref{eq_single_particle}) is due to single-particle interference in one of the MZIs. Single-particle interference appears, when the wavefunctions traveling through the upper and the lower arm of the interferometer have an overlap at the interferometer exit, which is only possible if the path difference of the interferometer arms is at most of the order of the usual first-order coherence length. Thus this interference term is suppressed as soon as the difference in the traversal times of the respective interferometer $\beta=\mathrm{L,R}$, given by $\Delta\tau_{\beta} = \tau^\mathrm{ u}_{ \beta\alpha}-\tau^\mathrm{d}_{\beta\alpha}$ is large compared to the half width $\Gamma_\alpha$ of a current pulse emitted by the source $\alpha=A,B$.  The suppression of the single-particle interference can therefore be used as a measure for the spreading of the wavepacket emitted by a driven capacitor. The path difference is tunable in experiment by  side gates applied to the edge states \cite{MZI}. 
We are now interested in the situation where the flux-dependence of the currents vanishes, and interference effects can be attributed to two-particle correlations.
\\ \indent
\textit{The noise.}---
Two-particle correlations can be observed in the noise properties.
We calculate the symmetrized zero-frequency  noise power (shot noise) \cite{BB00} for
currents flowing into contacts $1$ and $2$.
The equation for noise complementary to Eq.\,(\ref{eq_current_general}) reads,
\begin{eqnarray}
&&\mathcal{P}_{12} = \frac{e^2}{h}   \sum\limits_{n
= -\infty}^{\infty} \mathrm{sign}(n) \int\limits_{\mu- n\hbar\Omega}^{\mu}
\hspace*{-0.3cm} dE \int\limits_{0}^{\cal T} \frac{dt}{{\cal
T}}\int\limits_{0}^{\cal T} \frac{dt^\prime }{{\cal
T}} e^{in\Omega (t - t^{\prime})} \nonumber \\
& &\times \sum\limits_{l, j = 1}^{4} S^{}_{1l}(t,E) S_{1j}^{*}(t,E_{n}) S_{2l}^{*}(t^\prime,E) S^{}_{2j}(t^\prime,E_{n}) \,, \label{noise_gen}
\end{eqnarray}
where  $E_{n} =
E+n\hbar\Omega$.
If the arm length differences of the
MZIs are commensurate, $\Delta\tau:=\Delta\tau_\mathrm{L}=\mp\Delta\tau_\mathrm{R}$,
we find in the zeroth order in   $\Omega\Gamma_{\alpha}$,
\begin{eqnarray}\label{eq_shot_noise}
&& \mathcal{P}_{12}  =  - \mathcal{P}_{0} \sum_{s=\mathrm{e,h}} \big\{
T_\mathrm{L} T_\mathrm{R} \left[ 1 - L(\Delta
t^s) \right]\\
&& - \gamma_\mathrm{L} \gamma_\mathrm{R} \cos\left( \Phi_\mathrm{L} \pm
\Phi_\mathrm{R} \right) \left[L(\Delta t^s - \Delta\tau) + L(\Delta t^s +
\Delta\tau)\right]\big\}, \nonumber
\end{eqnarray}
with the classical MZI transmission probability  $T_\beta=T^\mathrm{l}_\beta T^\mathrm{r}_\beta+R^\mathrm{l}_\beta R^\mathrm{r}_\beta$. 
Here $\Delta t^s = t^s_{A} - t^s_{B} + \Delta\tau_{AB}$ depends on the difference of emission times $t^s_{\alpha}$ and the time delay $\Delta\tau_{AB}=\tau^\mathrm{u}_{\beta
A}-\tau^\mathrm{u}_{\beta B}$ due to the asymmetry of the set-up; the Lorentzians are defined by
$L(X) = 4\Gamma_{A}\Gamma_{B}/[X^2 + (\Gamma_{A} + \Gamma_{B})^2]$ and
$\mathcal{P}_{0} = (2e^2/\pi) T_{C} R_{C}
\Omega$ is (minus) the shot noise produced by the central QPC alone \cite{olkhovskaya08}.
\\ \indent
The second line in Eq.\,(\ref{eq_shot_noise}) is a magnetic flux-dependent contribution, appearing under two conditions.
First, the interferometers have to be commensurate, $\Delta\tau_\mathrm{L} = \mp\Delta\tau_\mathrm{R}$.
Second, the emission times of the
cavities are such that  a collision of two electrons (and/or two holes \cite{note1}) can take place at the interferometer outputs, $|\Delta t^{s} \pm \Delta\tau| \leq \Gamma_{\alpha}$.
Both conditions together imply that the collisions take place at \textit{both} interferometers. That results in an appearance of non-local two-particle correlations irrelevant for the current but with a pronounced effect in the noise.
In Fig.\,\ref{fig_noise} we show the shot noise for $\Delta\tau_\mathrm{L} = \Delta\tau_\mathrm{R}$, as a function of the magnetic flux difference and the phase shift $\varphi$ between potentials $U_{A}(t) = U_{A}\cos(\Omega t)$ and $U_{B}(t) = U_{B}\cos(\Omega t + \varphi)$ acting onto the capacitors $A$ and $B$.
We choose the difference between electron and hole emission times from the two sources to be equal. Each source emits one electron and one hole during the period ${\cal T}$.
Varying the phase $\varphi$ we change the time $t^e_{B}$ when the capacitor $B$ emits an electron.
At $\varphi = \varphi_{0}$ the condition $\Delta t^e - \Delta\tau=0$ is satisfied and an electron emitted by the capacitor $A$ and moving along the lower arm of an interferometer can collide (overlap) with an electron emitted by the capacitor $B$ and moving along the upper arm of the same interferometer (vice versa for  $\varphi = -\varphi_{0}$). 
Therefore, a mere variation of the phase difference between the driving potentials can switch on or switch off the two-particle AB-effect.
Note the dip at $\varphi = 0$ is the fermionic HOM effect \cite{olkhovskaya08}.
\\ \indent
\textit{Entanglement.}---
Now we show that these correlations are quantum, i.e., the emitted electrons are orbitally entangled in pairs.
The same, of course, is valid with respect to holes.
The degree of entanglement of the two-particle state, expected whenever the shot-noise of the system becomes flux dependent, can be measured in terms of the concurrence \cite{HW97_concurrence}. A wavepacket created at the source $A$  by the operators $\hat{A}^\dagger=  \int dk \ w(k)e^{ikv_\mathrm{F}(t_A-t)}\hat{a}^\dagger(k)$
 acting on the Fermi sea $|0\rangle$ is scattered into
 \begin{eqnarray}
 && \hat{A}^\dagger_\mathrm{out}(t) =  \int dk \ w(k)\left[
t_C t_{R}^{l}\mathcal{F}_\mathrm{R}^\mathrm{u}\hat{g}_{\mathrm{R}2}^\dagger(k)\right.\\
&&\left.
+t_Cr_{R}^{l}\mathcal{F}_\mathrm{R}^\mathrm{d}\hat{g}_{\mathrm{R}1}^\dagger(k)+ r_Ct_{L}^{l}\mathcal{F}_\mathrm{L}^\mathrm{u}\hat{g}_{\mathrm{L}2}^\dagger(k)+r_Cr_{L}^{l}\mathcal{F}_\mathrm{L}^\mathrm{d}\hat{g}_{\mathrm{L}1}^\dagger(k)
\right],\nonumber
 \end{eqnarray}
where $\mathcal{F}_\beta^\mathrm{u/d}=e^{i\bar{\Phi}_{A\beta}^\mathrm{u/d}}e^{ikv_\mathrm{F}(t_A+\tau_{A\beta}^\mathrm{u/d}-t)}$, (analogously for $\hat{B}^\dagger$ and $\hat{B}_\mathrm{out}^\dagger(t)$; for simplicity, we suppose that the wavepackets have the same shape, $w(k)$). The creation operators $\hat{g}(k)$ are related to the wavepacket detected at the four interferometer outcomes, representing the states $|1\rangle_\mathrm{L},\ |2\rangle_\mathrm{L},\ |1\rangle_\mathrm{R},\ |2\rangle_\mathrm{R}$ (see Fig.\,\ref{fig_collision}). 
We consider a part of the outgoing state corresponding to events with one particle to the left and one particle to the right.
Decomposing this outgoing two-particle state, 
 $|\Psi_\mathrm{out}\rangle=\hat{A}_\mathrm{out}^\dagger(t)\hat{B}_\mathrm{out}^\dagger(t)|0\rangle=\sum_{n,m=1}^2\chi_{nm}|n\rangle_\mathrm{L}|m\rangle_\mathrm{R}$, we
 choose the two observation times, $t_\mathrm{L}$ and $t_\mathrm{R}$ 
 such that each basis state consists of a product of single particle states detected at the left and at the right side of the system. 
 The concurrence is calculated from $\mathcal{C}=2\sqrt{\mathrm{det}(\chi\chi^\dagger)}$ \cite{beenakker06}. Whenever the collision conditions leading to a flux dependence of the noise are \textit{not} fulfilled at least three elements of $\chi$ vanish and the concurrence is therefore zero, $\mathcal{C}=0$. We consider now the time differences of the interferometer paths to be equal for the two interferometers and choose the times fulfilling the collision conditions as measuring times, $t_\mathrm{L}=t_{B}+\tau_{B\mathrm{L}}^\mathrm{d}=t_{A}+\tau_{A\mathrm{L}}^\mathrm{u}$ and $t_\mathrm{R}=t_{A}+\tau_{A\mathrm{R}}^\mathrm{u}=t_{B}+\tau_{B\mathrm{R}}^\mathrm{d}$. The concurrence is then 
 \begin{equation}
 \mathcal{C}=2T_CR_C\sqrt{
T_\mathrm{R}^{l}R_\mathrm{L}^{l} 
T_\mathrm{L}^{l}R_\mathrm{R}^{l}}/
(
T_C^2T_\mathrm{R}^{l}R_\mathrm{L}^{l} + 
R_C^2T_\mathrm{L}^{l}R_\mathrm{R}^{l}).
 \end{equation}
 Full entanglement, $\mathcal{C}=1$, is obtained by tuning the capacitors and choosing the appropriate measuring times, when the transmission and reflection probabilities are all equal. 
\\ \indent
To witness the entanglement found above, which is produced in definite bins of time, we use the violation of a Bell inequality (BI). 
Following Glauber \cite{Glauber63} we introduce the joint probability ${\cal N}_{12}$ to detect one electron (a wave packet) at the contact $1$ at the time of collision $t_\mathrm{L}$ and the other electron at the contact $2$ at the time of collision $t_\mathrm{R}$.
For simplicity we make no distinction between the time when a particle passed the interferometer output and the time of detection.  
This quantity can be calculated as follows, ${\cal N}_{12} = \delta {\cal N}_{12} + {\cal N}_{1} {\cal N}_{2}$, where ${\cal N}_{j}$ is the mean number of electrons detected at the contact $j = 1, 2$ at the corresponding time, and $\delta {\cal N}_{12}$ is a correlator. 
To calculate, e.g., $e{\cal N}_{1}$ we integrate the current $I_{1}(t)$, Eq.\,(\ref{eq_current_general}), near $t_\mathrm{L}$ over a time interval $\tau_{m}$ longer than the width of a current pulse $\Gamma_{\alpha}$ but shorter than the distance between different pulses. 
We find, ${\cal N}_{1} = R_C T^\mathrm{l}_\mathrm{L} T^\mathrm{r}_\mathrm{L} + T_C R^\mathrm{l}_\mathrm{L} R^\mathrm{r}_\mathrm{L}$ and $ {\cal N}_{2} = T_C T^\mathrm{l}_\mathrm{R} T^\mathrm{r}_\mathrm{R} + R_C R^\mathrm{l}_\mathrm{R} R^\mathrm{r}_\mathrm{R}$.
To calculate $\delta {\cal N}_{12}$, which is proportional to the shot noise, we modify Eq.\,(\ref{noise_gen}) in the following way.
The total shot noise ${\cal P}_{12}$ comprises contributions from particles arriving at contacts at any time during the driving period ${\cal T}$. 
Since we are interested in the contribution of electrons, we restrict the integral over $t$ ($t^{\prime}$) to the interval $\tau_{m}$ around $t_\mathrm{L}$ ($t_\mathrm{R}$).  
The time-resolved shot noise ${\cal P}_{12}(t_\mathrm{L},t_\mathrm{R})$ obtained thus defines $\delta{\cal N}_{12} = (\pi/\Omega) {\cal P}_{12}(t_\mathrm{L},t_\mathrm{R})/e^2$.
Calculations yield, $\delta {\cal N}_{12} = \delta {\cal N}_{12}^{A} + \delta {\cal N}_{12}^{B} + \delta {\cal N}_{12}^{AB}$, where $\delta {\cal N}_{12}^{A} = - R_C T_C T^\mathrm{l}_\mathrm{L} T^\mathrm{r}_\mathrm{L} T^\mathrm{l}_\mathrm{R} T^\mathrm{r}_\mathrm{R}$ and $\delta {\cal N}_{12}^{B} = - R_C T_C R^\mathrm{l}_\mathrm{L} R^\mathrm{r}_\mathrm{L} R^\mathrm{l}_\mathrm{R} R^\mathrm{r}_\mathrm{R}$ are due to the capacitors $A$ and $B$ alone, and the correlation contribution,
\begin{equation}
\delta {\cal N}_{12}^{AB} = 2 R_C T_C \gamma_\mathrm{L} \gamma_\mathrm{R} L(\Delta t^{e} - \Delta\tau) \cos\left( \Phi_\mathrm{L} - \Phi_\mathrm{R} \right)\,, \label{eq_correlator} 
\end{equation}
\noindent
depends on the wave-packet overlap $L(\Delta t^{e} - \Delta\tau)$.
\\ \indent
To test a BI \cite{Bell65,CHSH69} we use the four joint probabilities to detect one electron at the contacts $1$ (or $3$) at the time $t_\mathrm{L}$ and another electron at the contacts $2$ (or $4$) at time $t_\mathrm{R}$.
For the normalized correlation function, 
$E = \left( {\cal N}_{12} + {\cal N}_{34} - {\cal N}_{14} - {\cal N }_{ 32} \right)/ \left( {\cal N}_{12} + {\cal N}_{34} + {\cal N}_{14} + {\cal N }_{ 32} \right)$, we find, $E = L(\Delta t^{e} - \Delta\tau) \cos\left( \Phi_\mathrm{L} - \Phi_\mathrm{R} \right)$, if the transmissions at all the QPCs are $1/2$.
The magnetic fluxes $\Phi_\mathrm{L(R)}$ have no effect on the collision times $t_\mathrm{L(R)}$. Therefore, we can choose four sets of $\Phi_\mathrm{L} - \Phi_\mathrm{R}$, obtaining four different values of $E$, see, e.g., Ref.\,\cite{CHSH69}, to maximally violate a BI. 
This inequality holds for $L(\Delta t^{e} - \Delta\tau) > 1/\sqrt{2}$, i.e. even at partial overlap of wave-packets: for $\Gamma : = \Gamma_{A} = \Gamma_{B}$, we find the condition $\left|\Delta t^{e} - \Delta\tau\right| \lesssim 1.2\Gamma $.
\\ \indent
\textit{Conclusion.}--- 
Two uncorrelated but synchronized single-particle emitters can produce entangled pairs of electrons (holes). The orbital entanglement exists in bins of time with a well defined position within the driving period.  Its key signature is a non-local Aharonov-Bohm effect in distant loops. Simply changing the phase difference between the potentials driving the capacitors switches the entanglement on or off in a controlled manner. A successful realization of our proposal would open up new perspectives for a coherent quantum electronics. 
\\ \indent
We thank P. Samuelsson for useful discussion. 
We acknowledge the support of the Swiss NSF, the program for MANEP, and the EU project SUBTLE.



\begin{thebibliography}{99}

\bibitem{hanburybrown56}   R. Hanbury Brown and R. Q. Twiss, Nature (London) {\bf 178}, 1046 (1956).
 
 
\bibitem{hong87}       C. K. Hong, Z. Y. Ou, and L. Mandel, Phys. Rev. Lett. {\bf 59}, 2044 (1987).



\bibitem{buttiker91}      M. B\"uttiker,  Physica B (Amsterdam) {\bf 175}, 199 (1991); 
                                         Phys. Rev. Lett. {\bf 68}, 843 (1992).  


\bibitem{feve07}  
                     G. F\`eve, \textit{et al.},
                     Science {\bf 316}, 1169 (2007). 


\bibitem{olkhovskaya08} S. Ol'khovskaya,  \textit{et al.},
                      Phys. Rev. Lett. {\bf 101}, 166802 (2008).


\bibitem{franson89} J. D. Franson,  Phys. Rev. Lett. {\bf 62}, 2205 (1989).


\bibitem{aharonov59}  Y. Aharonov and D. Bohm, Phys. Rev. {\bf 115}, 485 (1959).


\bibitem{Bell65}   J. S. Bell, Rev. Mod. Phys. {\bf 38}, 447 (1966).


\bibitem{samuelsson} P. Samuelsson, E. V. Sukhorukov, and M. B\"uttiker, Phys. Rev. Lett. {\bf 92},   
                          026805 (2004);
                          P. Samuelsson, I. Neder, and M. B\"{u}ttiker, Phys. Rev. Lett. {\bf 102}, 106804  (2009).
              
              
\bibitem{neder07}  I. Neder, \textit{et al.},
		    Nature {\bf 448}, 333 (2007).


\bibitem{yurke92} B. Yurke and D. Stoler, Phys. Rev. A {\bf 46}, 2229 (1992).
 
 
\bibitem{entpump}   P. Samuelsson and M. B\"{u}ttiker,  Phys. Rev. B {\bf 71}, 245317 (2005);
                                    C. W. J. Beenakker, M. Titov, and B. Trauzettel,  Phys. Rev. Lett. {\bf 94}, 186804
                                    (2005).


\bibitem{MZI}  Y. Ji, \textit{et al.},  
                         Nature {\bf 422}, 415  (2003);
                         P. Roulleau, \textit{et al.}
                          Phys. Rev. Lett. {\bf 100}, 126802 (2008).


\bibitem{chung}  V. S.-W. Chung, P. Samuelsson, and M. B\"uttiker, Phys. Rev. B {\bf 72}, 125320
                        (2005);  
                         S.-W. V.  Chung, M. Moskalets, and P. Samuelsson, Phys. Rev. B  {\bf 75}, 115332 (2007).


\bibitem{moskalets08}  M. Moskalets, P. Samuelsson, and M. B\"uttiker, Phys. Rev. Lett. {\bf 100}, 086601    
                         (2008).


\bibitem{KSL08}  J. Keeling, A. V. Shytov, and L. S. Levitov, Phys. Rev. Lett. {\bf 101}, 196404 (2008).


\bibitem{splett08} J. Splettstoesser,  \textit{et al.},
                         Phys. Rev. B {\bf 78}, 205110 (2008).
 
 
\bibitem{BB00}  Ya. M. Blanter and M. B\"uttiker, Phys. Rep. {\bf 336}, 1 (2000).


\bibitem{note1}  The electron-hole collisions have no effect onto the noise.


\bibitem{HW97_concurrence}   W. K. Wootters, Phys. Rev. Lett. {\bf 80}, 2245 (1998).

\bibitem{beenakker06}  C. W. J. Beenakker, \textit{Proc. Int. School Phys.}  "E. Fermi", Vol. 162 (IOS  
                      Press,  Amsterdam, 2006).


\bibitem{Glauber63} G. J. Glauber, Phys. Rev. {\bf 130}, 2529 (1963).

\bibitem{CHSH69}  J. F. Clauser, M. A. Horne, A. Shimony, and R. A. Holt, Phys. Rev. Lett. {\bf 23}, 880 
                               (1969);
                               J. F. Clauser and A. Shimony, Rep. Progr. Phys. {\bf 41}, 1881 (1978).


\end{thebibliography}
\end{document}